\documentstyle[12pt,aaspp]{article}
\begin{document}

\title{ARE THE OGLE MICROLENSES IN THE GALACTIC BAR?}

\author{B.~Paczy\'nski\altaffilmark{1}, K.~Z.~Stanek\altaffilmark{1,5},
A.~Udalski\altaffilmark{2,6}, M.~Szyma\'nski\altaffilmark{2,6}}
\affil{\tt e-mail I: (bp,stanek)@astro.princeton.edu,
(udalski,msz)@sirius.astrouw.edu.pl}

\author{and}

\author{J.~Ka\l u\.zny\altaffilmark{2,6}, M.~Kubiak\altaffilmark{2,6},
M.~Mateo\altaffilmark{3} and W.~Krzemi\'nski\altaffilmark{4}}
\affil{\tt e-mail I: (jka,mk)@sirius.astrouw.edu.pl,
mateo@astro.lsa.umich.edu, wojtek@roses.ctio.noao.edu}

\altaffiltext{1}{Princeton University Observatory, Princeton, NJ 08544--1001}
\altaffiltext{2}{Warsaw University Observatory, Al.~Ujazdowskie 4,
00--478 Warszawa, Poland}
\altaffiltext{3}{Department of Astronomy, University of Michigan, 821 Dennison
Bldg., Ann Arbor, MI 48109--1090}
\altaffiltext{4}{Carnegie Observatories, Las Campanas Observatory, Casilla
601, La Serena, Chile}
\altaffiltext{5}{On leave from N.~Copernicus Astronomical Center,
Bartycka 18, Warszawa 00--716, Poland}
\altaffiltext{6}{Visiting Astronomer, Princeton University Observatory,
Princeton, NJ 08544--1001}

\centerline{\it Received: 1994 .........................................}

\begin{abstract}

The analysis of the first two years of OGLE data revealed 9 microlensing
events of the galactic bulge stars, with the characteristic
time scales in the range $ 8.6 < t_0 < 62 $ days, where
$ t_0 = R_E / V $. The optical depth to microlensing is
larger than $ ( 3.3 \pm 1.2 ) \times 10^{-6}$, in excess of current
theoretical estimates, indicating a much higher efficiency for
microlensing by either bulge or disk lenses. We argue that the
lenses are likely to be ordinary stars in the galactic bar, which has
its long axis elongated towards us. A relation between $ t_0 $
and the lens masses remains unknown until a quantitative model
of bar microlensing becomes available. At this time we have no
evidence that the OGLE events are related to dark matter.

The geometry of lens distribution
can be determined observationally when the microlensing rate
is measured over a larger range of galactic longitudes,
like $ -10^o < l < +10^o $, and the relative proper motions of
the galactic bulge (bar) stars are measured with the HST.

\end{abstract}


\section{INTRODUCTION}

The OGLE project (Optical Gravitational Lensing Experiment)
was described by Udalski et al.~(1992), and the
lensing events were reported by Udalski et al.~(1993, 1994a,b). All
observations were done with the 1 meter Swope telescope at the Las
Campanas Observatory, operated by the Carnegie Institution of Washington.
The detector was a single Loral CCD with $(2k)^2$ pixels. We used
a somewhat modified DoPhot photometric software (Schechter, Mateo \&
Saha 1993) to extract stellar magnitudes from the CCD frames.
All technical details were provided in Udalski et al.~(1992, 1994b)
and references therein. The location of all 9 OGLE events
in the color-magnitude diagram is shown in Fig. 1.

The first estimate of the optical depth to gravitational microlensing
of the galactic bulge stars as observed by the OGLE has recently
been published (Udalski et al.~1994b). The result is surprising:
the minimum optical depth is estimated to be $ \sim ( 3.3 \pm 1.2 )
\times 10^{-6}$ in Baade's Window and the nearby galactic bar fields.
Theoretical
models predicted a much lower value, somewhere in the range of
$ 0.5 - 1.0 \times 10^{-6}$ (Paczy\'nski 1991; Griest et al.~1991;
Kiraga and Paczy\'nski 1994, hereinafter referred to as KP;
Giudice et al.~1994). The aim of this paper is to discuss
this apparent discrepancy and to propose a simple observational
test that could definitely locate the apparent excess of lensing
masses either in the disk or in the bulge.
Future observations will
resolve this uncertainty. Here, we adopt the OGLE estimate as correct,
and we follow the consequences. It is important to note that
a similar (high) estimate of the optical depth to the galactic bulge stars
has been recently obtained by the MACHO collaboration (Alcock et al.~1994).

The 9 OGLE events satisfied a variety of distribution tests expected
of gravitational microlensing (Udalski et al.~1994b), but we cannot claim
that we have proven beyond any reasonable doubt that they are indeed
due to lensing. Nevertheless, the case for lensing is strong, as
no other kind of intrinsically variable stars of this type is known.
The lensed objects are scattered over a large region of the color-magnitude
diagram, in proportion to the number density of the observed stars as
apparent in Fig. 1 and quantitatively assessed by Udalski et al.~(1994b).
In this large region of the diagram, covering roughly solar type
stars at the main sequence turn-off point and red subgiants up to
the ``red clump'', no intrinsically variable stars are know, at least
not with amplitudes as large as our lens candidates.
Therefore, throughout this paper we assume that the 9 OGLE events
were due to gravitational microlensing. We adopt the traditional
definition of the event time scale: $ t_0 = R_E/V $ (Paczy\'nski 1986),
where $ R_E $ is the Einstein ring radius, and $ V $ is the relative
velocity of the source, lens, and observer.

\section{MODELS}

We follow KP in our model of the
galactic disk and the galactic bulge contribution to the lensing.
We ignore the possible contribution of the dark galactic halo
as the OGLE was aimed at the event time scale characteristic of
ordinary stellar masses rather than very low or very high mass
dark objects. In any case, the optical depth in the direction of
Baade's Window due to halo objects
of any kind is only $ \sim 0.13 \times 10^{-6} $ (Griest et al.~1991),
much less than the contribution due to any disk or bulge models.

It is a convenient coincidence that while looking
at the galactic latitude $ |b| = 4^o $ the radial and the vertical
disk exponentials (Bahcall 1986) almost exactly cancel each other if the
galactic longitude is small, $ |l| < 20^o $ or so, which is the
case for all our fields (cf.~Paczy\'nski 1991, Udalski et al.~1994b).
This implies that the number density of stars is expected to be
approximately constant along the line of sight. However, this conventional
model is in apparent conflict with the observations.

First, the recent measurements of the disk scale height by Kent et al.
(1991) indicate that the scale height decreases towards the galactic
center, which implies that the number density of disk stars as seen
through Baade's Window should gradually decrease with the distance
from observer. Second, preliminary
analysis of the color -- magnitude diagrams obtained with the
OGLE revealed an unexpected feature in the distribution of disk
stars (Paczy\'nski et al.~1994): the density is {\it observed} to be
uniform out to some distance $ d_{max} \approx 3-4 ~ kpc $
but it decreases by a large factor beyond $ d_{max}$, as if the disk
was nearly empty within radius $ r_{in} = R_0 - d_{max}$, where
$ R_0 $ is our distance from the galactic center.
The fact that the inner disk has low density has been noticed
before with the OH/IR stars (Baud et al. 1981, Blommaert et al. 1994).
Unfortunately, the current knowledge of the density distribution
is so limited that for the
purpose of this paper we adopt a simple distributions of
the stellar density along the line of sight through Baade's Window:
\begin{equation}
\rho (d) = \rho_0 ~~~~~ { \rm for } ~~~~ d \leq d_{max} ~ , ~~~~~~~~~~
\rho (d) = 0 ~~~~~ { \rm for } ~~~~ d > d_{max},
\end{equation}
where $ \rho_0 $ is the local disk density near the sun.

Following KP we adopt Kent's (1992)
bulge model. The lensing is made possible because the bulge
has a finite radial extent, with the stars in front lensing those
in the back. We adopt the bulge
cumulative luminosity function to be a power law with the slope (--2).
The results depend only very weakly on this number (KP).
The main limitation of the adopted structure
is its axial symmetry, while there is plenty of evidence
that the bulge is in fact bar-like (cf.~de Vaucouleurs 1964,
Binney et al.~1991, Blitz \& Spergel 1991, Blitz 1993, Sellwood 1993,
Stanek et al.~1994, and references therein).

Following KP we adopt the same power law
mass function for the bulge and for the disk stars,
with the number density proportional to $ M^{-2}$,
i.e.~equal amount of mass per logarithmic mass interval.

We consider a conventional scenario first. We adopt the
mass density in ordinary stars near the sun to be $ \rho_0 = 0.05 ~
M_{\odot} ~ pc^{-3}$ (Bahcall \& Soneira 1980), with the mass range
from $ 0.1 M_{\odot}$ to $ 1 M_{\odot}$, and a constant number
density all the way to the galactic bulge, i.e. $ d_{max} = 8 ~ kpc $.
We adopt the theoretical distribution of microlensing event rate
as a function of event time scale $ \Gamma (t_0) $
given by KP and combine it
with the OGLE efficiency for event detection $ \epsilon (t_0) $
as given by Udalski et al.~(1994b)\footnotemark[1],
\footnotetext[1]{
The OGLE efficiency averaged over the 1992 and 1993 observing seasons
was 0.0015, 0.029, 0.11, 0.22, 0.26 for $ t_0 = $ 1.0, 3.2, 10.0, 31.6,
and 100.0 days, respectively.}
to obtain the cumulative number of events with the duration
less than $ t_0 $:
\begin{equation}
N(\leq t_0 ) = \int _0^{t_0} \Gamma (t_0) \epsilon (t_0) ~ dt_0 ~~ .
\end{equation}
The total number of events with $ t_0 \leq 100 $ days expected in this model
is 2.2, four times less than observed.

Now we modify our model by increasing the local disk density
by a factor of 3 to $ \rho_0 = 0.15 ~ M_{\odot} ~ pc^{-3}$,
i.e.~somewhat in excess of the estimate of the total local disk mass
by Bahcall \& Soneira (1980), the estimate considered too high by many
recent investigations (Kuijken \& Gilmore 1991, and references therein).
In order not to be in
direct conflict with what is observed near the sun we extend the range
of lens masses into the brown dwarf region, i.e.~we adopt a
mass function extending over two decades: $ 0.01 \leq
M/M_{\odot} \leq 1.0 $. With the adopted power law there is just
as much mass in objects below $ 0.1 ~ M_{\odot} $
(brown dwarfs) as in objects above $ 0.1 ~ M_{\odot} $ (stars).
The model predicts 4.2 events, still a factor two short of the OGLE 9.

We may also consider just the optical depth to microlensing.
The galactic bulge microlensing model of
KP contributes only $ 0.5 \times 10^{-6} $
to the overall optical depth to gravitational microlensing,
the halo contributes only $ 0.13 \times 10^{-6} $ (Griest et al.~1991),
leaving $ \tau_{disk} = 2.7 \times 10^{-6} $ to the disk
if the total is to agree with the OGLE result.
A simple formula relates the disk optical depth
to the local disk mass density (cf.~eq.~1 of KP):
\begin{equation}
\tau_{disk} = 6.5 \times 10^{-6} ( 3x^2 - 2x^3)
\left( { \rho_0 \over 1 ~ M_{\odot} ~ pc^{-3} } \right) ~~ ,
{}~~~~~~~ x \equiv d_{max} / R_0 ~ ,
\end{equation}
where all symbols have the same meaning as in eq.~(1).
Adopting $ \tau _{disk} = 2.7 \times 10^{-6} $ implies that
$ \rho _0 = 0.415 ~ M_{\odot} ~ pc^{-3} $ for the ``full'' disk model
($x=1$), and $ \rho _0 = 0.83 ~ M_{\odot} ~ pc^{-3} $ for the ``hollow''
disk model ($x=0.5$).
These densities are a factor of 3 or 6 higher
than the highest dynamical estimate (Kuijken \& Gilmore 1991; Bahcall,
Flynn \& Gould 1992). This discrepancy firmly rules out
the disk as the main site of OGLE lenses.

Nevertheless, it is interesting to ask a question: what would be the
lens masses if they were in a super-dense disk? It is very tempting
to do the exercise, as OGLE provides the first information about the
distribution of event time scales, $ t_0 $. We adopt the following
procedure, which may be used for any model of the lens distribution.
We take the rate of events as a function of event time scale
$ \Gamma (t_0) $ from the model. We take the efficiency of lens
detection $ \epsilon (t_0) $ from the experiment. We calculate the
expected probability that an observed event should have a duration
less than $ t_0 $:
\begin{equation}
P(\leq t_0 ) =
\left( \int _0^{t_0} \Gamma (t_0) \epsilon (t_0) ~ dt_0 \right)
\left( \int _0^{\infty} \Gamma (t_0) \epsilon (t_0) ~ dt_0 \right) ^{-1} .
\end{equation}
If the model is correct we expect the values of integral
probability $ P_k $ for all events to be uniformly but randomly
distributed in the interval (0,1).

We adopted a ``full'' disk model ($ d_{max} = 8 ~ kpc $)
with all lenses having the same mass $M$.
We varied $M$ in small logarithmic
steps over a large range, and we calculated the
probability $ P_k $ for all nine OGLE lenses for all the models.
The distribution of those is shown in Fig.~2 for
$ -1 \leq \log ~ (M/M_{\odot}) \leq 1 $.
The small points indicate values of $ P_k $ for the nine OGLE
lenses. They cluster near $ P=0 $ if the lenses are massive
and near $ P=1 $ if the lenses are light. For a random but uniform
distribution we require $ \langle P \rangle = 0.5 $.
This is achieved for the lens mass $ M/M_{\odot} = 0.65 $,
which is a typical stellar mass. Such objects would be
very difficult to hide. Therefore, even if the dynamical
mass estimates for the local disk density (cf. Kuijken \& Gilmore 1991;
Bahcall, Flynn \& Gould 1992)
were wrong for some obscure reason, we would be still faced with the
problem how to make the numerous $ 0.65 ~ M_{\odot} $ objects
escape the detection in the local disk.

One might claim that perhaps what we know about the local disk does
not apply to the disk half way between us and the bulge. For example,
if the radial scale length is as short as $ 2.5 ~ kpc $ then
the radial increase in the density of stars would outweigh the
decrease of density due to our line of sight getting out of the plane.
Or the disk could be thicker some distance towards
the galactic center. However, there is no observational evidence
to support such hypothetical claims, and in fact the observational
evidence is to the contrary: the {\it observed} number density of
stars along the line of sight through Baade's Window appears to be
uniform between the sun and $ d_{max} \approx 3-4 ~ kpc $, and it
declines dramatically beyond $ d_{max} $ (Paczy\'nski et al.~1994).

\section{DISCUSSION}

In the previous section we have demonstrated that the rate of OGLE
events cannot be explained by any reasonable galactic disk model. The
rate cannot be explained by the only available galactic bulge model either.
So, what is going on? We can only speculate at this time,
but we think that a good case can be made for the lenses to be in
the galactic bar, i.e.~highly non-axially symmetric galactic bulge.
There was evidence for the presence of a bar in the inner galaxy for
about three decades (de Vaucouleurs 1964), but only recently it became
popular (cf.~Blitz 1993 and references therein).
It is most clearly seen in the distribution of disk globular
clusters projected onto the galactic plane, as shown in Fig.~5 of
Blitz (1993). The bar seems to have its long axis inclined by $ \sim 15^o $
to our line of sight. This orientation of the bar was also deduced
by Binney et al.~(1991) using the information about the observed gas
velocities. The presence of the bar is clearly detected with the
OGLE data as a difference of $ \sim 0.37 ~ mag $ in the apparent
magnitude of the ``red clump'' stars in the two opposite OGLE galactic
bar fields (Stanek et al.~1994).

The radial depth of the bulge/bar, $ \Delta d_b $, is much smaller
than its distance $ d_b \approx 8 ~ kpc $, hence its optical depth
to microlensing scales as $ \tau _b \sim \Sigma _b \Delta d_b $, where
$ \Sigma _b $ is the column mass density of the bulge/bar (cf. eq. 2 of KP).
In the KP model the bulge was assumed to be axisymmetric, with $ \Delta d_b
\approx d_b \Delta \varphi _b $, where $ \Delta \varphi _b $ is the
apparent angular extent of the bulge/bar.  If the bar axis ratio is
large, i.e. if $ f \equiv \Delta d_b / ( d_b \Delta \varphi _b ) \gg 1 $,
then its optical depth is increased by the factor $f$, and even more
if its column mass density $ \Sigma _b $ is increased.

There is another very important consequence of the bar-like
shape of the galactic bulge:
the average distance between the lens and the source is larger in
a bar and hence the Einstein ring radius $ R_E $ is larger.
%  The time scale of microlensing events is given as $ t_0 = R_E/V $.
In a bar we expect $ R_E $ to increase in comparison with the KP model,
leading to the increase of $ t_0 $ for lenses of the same mass.
For a given time scale $ t_0 $ the lens mass scales as $ M_{lens}
\sim 1 / \Delta d_b \sim 1/f $, i.e. as the inverse of
the bar axis ratio.

Unfortunately, the bar axis ratio $f$ is not well
constrained either by data or by models at this time,
but the radial extent of the bar may be
much larger than its apparent transverse extent.
If that is the case, i.e. if $f \gg 1 $, then
the KP optical depth is increased by the factor $f$,
and the inferred masses are reduced by the same factor $f$.

It is worth noticing that the bulge or bar mass estimates by Kent (1992)
and by Binney et al.~(1991) were based on the dynamics, not on the
observed light, i.e.~they refer to the {\it total} mass, luminous and
dark.  The bar mass as deduced by Binney et al.~(1991) was somewhat
larger than the bulge mass estimated by Kent (1992) in his axially
symmetric model.  This also increases a bit the optical depth of a
bar as compared to a bulge.

Had we blindly used the KP model and followed
the procedure described in the previous section and displayed in
Fig.~2, we would find the masses of bulge lenses to be
$ \sim 1.1 ~ M_{\odot} $.  In a bar with the axis ratio $f \gg 1 $
the corresponding lens mass would be reduced by the factor $f$.
In any case, there is no need at this time to
seek help of the low mass objects (brown dwarf) for the lensing as
the problem is just the opposite: the observed time scales are too long.

There may be yet another effect which may affect the estimate of
the lens masses.
Stellar orbits within a bar are highly elongated, and
the radial velocity dispersion (which is observed) may be much larger than
the velocity dispersion transverse to the line of sight, as claimed by
Binney et al.~(1991).  It is the transverse velocity $V$ that is relevant
for lensing, and the reduction of $V$ leads to the increase of the time scale
$ t_0 $.  Unfortunately, the situation is
somewhat confusing as Binney's et al. (1991) expectation does not
seem to be supported by the data (Spaenhauer et al. 1992), and by the
other bar models (Sellwood 1993, Zhao et al. 1994), which all seem to be
consistent with the isotropic velocity dispersion.

The apparent excess in the observed number of microlensing events as
expressed in terms of the optical depth to microlensing:
$ (3.3 \pm 1.2) \times 10^{-6} $ (Udalski et al.~1994b) is only a lower
limit, as the OGLE project was not sensitive to events shorter
than $ \sim 5 $ days or longer than $ \sim 100 $ days.  The optical depth
due to bulge lenses of {\it all} masses is only $ \sim 0.5 \times 10^{-6} $
in the axially symmetric KP model, and any reasonable disk model contributes
no more than that. It is virtually certain that the disk and the bulge
are both important contributors to the observed microlensing.
Their relative contribution has to be established observationally.
What is needed is a map of optical
depth to gravitational microlensing as a function of galactic
coordinates. This will take a lot of observing and a lot of
data processing, but the technology has been proven to work, i.e.
the task is feasible.
The disk contribution is expected to be fairly constant for
$ |l| \leq 20^o $, while the bulge contribution should be greatly
reduced at $ |l| \ge 10^o $ as compared to $ l \approx 0^o $
(KP, Evans 1994).
It is also very important to establish the radial depth of the bulge/bar
observationally, in order to have as direct as possible determination
of the axis ratio.

\centerline{-----------------}

Beginning with the 1994 season the OGLE project is capable of near real time
data processing (Paczy\'nski 1994). The new computer system automatically
signals the events while they are on the rise, making it possible to carry out
photometric and/or spectroscopic follow-up observations. The observers who
would like to be notified about the on-going events should send their request
to A.~Udalski (udalski@sirius.astrouw.edu.pl).

The photometry of the OGLE microlensing events, their finding charts,
as well as a regularly updated OGLE status report, including more information
about the ``early warning system'', can be found over Internet from
``sirius.astrouw.edu.pl'' host (148.81.8.1), using the ``anonymous ftp''
service (directory ``ogle'', files ``README'', ``ogle.status'',
``early.warning''). The file ``ogle.status''
contains the latest news and references to all OGLE related
papers, and the PostScript files of some publications, including
Udalski et al.~(1994b). The OGLE results are also available over
``World Wide Web'': ``http://www.astrouw.edu.pl/''.

\acknowledgements{It is a great pleasure to acknowledge that the many of
$ \Gamma (t_0) $ distributions used in this paper were kindly provided
by M.~Kiraga. It is also a pleasure to acknowledge important comments
by D.~N.~Spergel.  One of us (BP) is most grateful to A. Omont and many
other astronomers at the Institut d'Astrophysique in Paris for their
hospitality.  This project was supported with the NSF grants AST
9216494 and AST 9216830 and Polish KBN grants No 2-1173-9101 and BST475/A/94.}

\newpage

\begin{figure}
\begin{center}
{\bf FIGURE CAPTIONS}
\end{center}

%Fig. 1
\caption{The location of the 9 OGLE lensing events (large circles)
are shown in the color -- magnitude diagram for Baade's Window
and the galactic bar fields combined. Only $ \sim 5\% $ of all
OGLE stars are shown.
The dashed line shows the limit of lenses detectability given
by the condition $I<19.5$.}

%Fig. 2
\caption{The distribution of $ P_k( \leq t_0 ) $ values is shown with the
dots for all 9 OGLE lensing events for a series of models with various
values of the lens masses, $ \log M $, with all lenses having the same
mass in a given model. The average value of the probability
$ \langle P \rangle $ is shown for every model with a filled triangle.
For the correct model the values of $ P_k $ should be randomly
distributed in the interval (0,1), and it should have
$ \langle P \rangle = 0.5 $. This model is shown
with a dashed horizontal line corresponding to $ M/M_{\odot} = 0.65 $.
The ``full disk'' with $ \rho _0 = 0.415 ~ M_{\odot} ~ pc^{-3} $ plus
the KP bulge contributed to the lensing.}
\end{figure}

\end{document}